\documentclass[preprint,12pt]{elsarticle}

\usepackage{amssymb}
\usepackage{xcolor}
\usepackage{rotating}
\usepackage{graphicx}
\usepackage{multirow}
\usepackage{graphicx,subcaption}
\usepackage{microtype}
\usepackage{array}
\usepackage{hyperref}
\begin{document}

\begin{frontmatter}

% Title
% \title{Neutron yield calculation with SOURCES4}
\title{Optimised neutron yield calculations from ($\alpha,n$) reactions with the modified SOURCES4 code}

\author{M. Parvu$^1$}
\author{P. Krawczun$^2$}
\author{V. A. Kudryavtsev$^2$}
\cortext[cor1]{corresponding author: M. Parvu, mihaela.parvu@unibuc.ro}

\address{$^1$Faculty of Physics, University of Bucharest, POBox 11, 077125, Magurele, Romania}
\address{$^2$Department of Physics and Astronomy, University of Sheffield, Sheffield, S3 7RH, United Kingdom}

\begin{abstract}
The sensitivity of underground experiments searching for rare events such as dark matter, neutrino interactions or several beyond the standard model phenomena is often limited by the background caused by neutrons from spontaneous fission and ($\alpha,n$) reactions. A number of codes exist to calculate neutron yields and energy spectra due to these processes. In this paper we present new calculations of neutron production using the modified SOURCES4 code with recently updated cross-sections for ($\alpha,n$) reactions and the comparison of the results with available experimental data. The cross-sections for ($\alpha,n$) reactions in SOURCES4 have been taken from reliable experimental data where possible, complemented by the calculations with EMPIRE 2.19/3.2.3, TALYS 1.96 or evaluated data library JENDL-5 where experimental data were scarce or unavailable. 
We present here our choice of the most reliable cross-sections for such calculations that match experimental data from alpha beams and radioactive decays.
\end{abstract}


\end{frontmatter}

\section{Introduction}
\label{intro}

Underground experiments for rare event searches in the low-energy ($\sim$keV-MeV-scale) region, such as neutrino interactions and phenomena beyond the standard model (SM) of particle physics are often limited by various backgrounds caused by radioactivity and cosmic rays. In underground experiments searching for dark matter WIMPs, 
%one of the most important background sources consists of
neutrons are one of the most dangerous background sources because they produce nuclear recoils 
%produced by these particles 
that are sometimes indistinguishable from those caused by dark matter candidates. 
%such as WIMPs. 
%Furthermore, 
In experiments focused on neutrinoless double beta decay ($0\nu\beta\beta$) $\gamma$-rays with energies of a few MeV,
produced via neutron inelastic scattering $(n, n’\gamma)$ or neutron capture $(n,\gamma)$ can give a signal in a range of interest for the search for the signal. 

Some background events can be reduced by rock overburden (cosmic rays), passive or active shielding (cosmic rays, neutrons and gammas from rock), additional veto systems (neutrons and gammas from detector components) and various discrimination techniques (multiple vertex events, electron recoil versus nuclear recoil events). Despite this, neutrons produced in spontaneous fission (SF) processes and ($\alpha,n$) reactions remain an important source of background in such experiments and their production rates and energy spectra need to be well understood so this type of background is under control.

In a recent paper \cite{whitepaper}, a broader range of applications for neutrons coming from $(\alpha,n)$ reactions is discussed, highlighting the importance of these studies.

There are several computer codes to calculate the neutron yields and energy spectra of these processes \cite{sources4,Gromov:2023iuh,neucbot,mendoza2019,usd,Vlaskin:2015hhf}. Spontaneous fission (SF) is well described by the parameterisation suggested in Ref.~\cite{watt} with parameters tuned to the measurements. The SF neutron yield does not depend on the material where the neutron emission occurs, but only on the concentration of $^{238}$U since other radioactive isotopes give negligible contribution from this process. The main problem is then the accurate calculation of ($\alpha,n$) reactions where alpha particles are produced in the decay chains of $^{235,238}$U and $^{232}$Th
%radioactive isotopes of uranium and thorium 
which can be found as impurities inside active media of detectors, in other detector components or in the environment. A comparison between different codes to calculate neutron yields and spectra and experimental data has been included in several papers, see for instance Refs.~\cite{neucbot,mendoza2019,sources4,fernandes,scorza,whitepaper}.

%The codes use cross sections of these reactions and transition probabilities to excited states as input together with energy losses of alphas as they travel through the material until they stop. 

Calculating neutron production from ($\alpha,n$) reactions in a material requires knowledge of the stopping powers for the alpha particles and the probability of neutron production, $P(E)$, as a function of the alpha particle energy, $E_{\alpha}$. The probability of producing a neutron, $P_i(E_{\alpha})$, from an alpha particle with energy $E_{\alpha}$, from material $i$, can be calculated 
%using the following expression 
as \cite{WILSON2009608}:

\begin{equation}
    P_i(E_{\alpha}) = \frac{N_i}{N} \int_0^{E_{\alpha}} \frac{\sigma_i (E)}{\varepsilon (E)} dE,
\end{equation}
where $N$ is the total atom density of the material $N_i$ is the atom density of nuclide $i$, $\varepsilon(E)=-\frac{1}{N} \frac{dE}{dx}$ is the stopping cross sections, $\sigma_i(E)$ is the microscopic cross section for ($\alpha,n$) reaction on nuclide $i$.

The SOURCES4 code can compute neutron production rates from ($\alpha, n$) reactions across four types of configurations: homogeneous, beam, interface, and three-region setups. It can also calculate neutron production rates from SF and delayed neutron emission. Neutron spectra are also calculated for each scenario.

%In this paper, we report on the improved calculations of neutron production with the SOURCES4 code which are essential for better understanding of the background in rare event searches that should lead to the increase of detector's sensitivity. 
In this paper, we present improved neutron production calculations using the SOURCES4 code, which are crucial for accurately characterising backgrounds in rare event searches and, ultimately, for evaluating and improving detector sensitivity. Recent developments of the code and libraries include the addition of recent experimental data for the cross sections and the optimisation of the choice of the cross-section for a particular isotope selecting it from experimental data where available or from a model where data are lacking. We have also increased the number of elements and isotopes that can be included in compounds in the SOURCES4 input file. We focus here on ($\alpha,n$) reactions caused by alphas from the radioactive decay chains of $^{238}$U, $^{235}$U and $^{232}$Th as the main contributors to the neutron background in low-background experiments. 

%This paper discusses the optimisation of cross-sections implemented in SOURCES4 and compares the calculated neutron yields with experimental data from alpha beams and the entire decay chain of U and Th. Additionally, a few other modifications to the SOURCES4 and TALYS codes are discussed.

With the newly implemented and optimised cross-sections, we have achieved more accurate neutron yield predictions for the majority of isotopes and compounds relevant to dark matter and neutrino experiments. This improvement is crucial for enhancing the reliability of background estimations in rare event searches, ultimately increasing the sensitivity of these experiments to new physics discoveries.

In Section~\ref{sec:sources4a} we briefly describe the SOURCES4 code and the cross-sections used. Section~\ref{Cross-sections-comp} includes the discussion of different sets of cross-sections for several isotopes. In Section~\ref{sec:comparison} the comparison of the SOURCES4 output with experimental data is presented for some materials. Conclusions are given in Section~\ref{sec:conclusions}.

\section{The SOURCES4 code to calculate neutron production in ($\alpha,n$) reactions}
\label{sec:sources4a}

The nuclear physics code SOURCES4 \cite{sources4} has been used for a long time in a number of applications. 
%It has several advantages over other similar techniques, such as flexible libraries of cross-sections and branching ratios, allowing for easy customisation and updates. 
The code library contains a range of cross-sections and branching ratios for each nuclide in a material with an option for a user to choose the most suitable dataset.
The code is capable of very fast calculations, providing, in a single run, total neutron spectra, spectra from interactions on individual isotopes, and spectra from various radioisotopes. It can also offer spectra from both the ground state and different excited states, making it versatile for different types of analysis. The code does not generate gamma rays from the de-excitation of final state nuclei and it cannot handle 'surface' contamination or related problems, which are common limitations among some similar codes.

The most recent version of the code is SOURCES4C \cite{sources4} but for historical reasons, we use the older version SOURCES4A \cite{sources4a}. Prior evaluations and release notes confirmed that the results obtained from both versions are the same or very similar when identical cross-sections and transition probabilities to the excited states are used in both versions. The code allows the user to select the total cross-section and branching ratios of the ($\alpha,n$) reaction for a specific isotope in a material. Additionally, the user has the capability to expand the library by adding more cross-sections and branching ratios.

The original code calculates neutron production for alpha energies up to 6.5~MeV and is not fully suitable for the evaluation of neutron rates from radioactive processes that involve alphas with energies up to about 9~MeV. The original code SOURCES4A was modified to remove the 6.5~MeV energy cut and an updated version now allows calculations of neutron production from alphas with energies up to about 10~MeV \cite{carson,tomasello2008}. Libraries of cross-sections and branching ratios were updated with calculated values from the TALYS 1.96 \cite{Koning:2023ixl}, and EMPIRE 2.19 and 3.2.3 \cite{Herman:2007} nuclear reaction codes, as well as JENDL-5 database \cite{JENDL5}, and extended to alpha energies up to 10 MeV (see \cite{carson,tomasello2008,lemrani,tomasello2010,tomasello-thesis} for early updates). Cross-sections and branching ratios were calculated with an 0.1~MeV step in alpha energy. Initial changes have been made to the version SOURCES4A and, since no changes in results were observed with the newer version of the code, SOURCES4A is continued to be used in most calculations. A comparison of cross-sections from EMPIRE 2.19 with some experimental data was published in Refs. \cite{tomasello2008,tomasello-thesis} and the results from the modified SOURCES4A code were used in a number of dark matter experiments (see, for example, Refs. \cite{eureca,edelweiss,lz,xenon1t}). We have recently modified the code to a) allow calculation of neutron yields and spectra from different materials in one run; b) allow the choice of the cross-section and branching ratios in the input file without the need to change positions of the cross-section or transition probabilities in the libraries (tape 3 and tape 4); c) increase the maximum number of target nuclides to 100; d) increase the maximum number of discrete nuclear levels to 500.

The user input for SOURCES4A includes either the energy of an alpha particle or the $Z$ and $A$ of the radioactive isotope (or several isotopes in the case of decay chains, for instance) with the number of radioactive atoms per unit volume. The user should also specify material composition (where the alpha sources are located) and isotopic composition for each element (only isotopes with cross-sections present in the code library can be included). 

An initial comparison of cross-sections from TALYS 1.9, EMPIRE 2.19/3.2.3, and experimental data was presented in Refs.~\cite{Kudryavtsev:2020eer,kudryavtsev2022an}. These papers also included some preliminary results on the comparison of neutron yields calculated using SOURCES4A with different cross-sections from TALYS 1.9, EMPIRE 2.19/3.2.3, and some experimental data. 

The TALYS and EMPIRE nuclear reaction codes based on statistical models provide an accurate description of reaction mechanisms over wide energy and nuclear mass ranges. These codes return reliable results for ($\alpha,n$) reactions within the energy range up to 200~MeV, for atomic numbers greater than 9 and mass numbers between 20 and 339. More precisely, 
%the codes are reasonably accurate for isotopes which do not have resonances in the energy dependent cross-sections. 
the codes are reasonably accurate for high-Z isotopes with energy thresholds above the resonance region. For light elements with a large number of resonances experimental data or evaluated data from the JENDL-5 library are to be used to achieve better accuracy. The alpha-particle sub-library in JENDL-5 provides evaluated cross-section data for alpha-induced reactions on 18 light isotopes from $^6$Li to $^{30}$Si, with an upper limit of 15~MeV for the incident energy. 

In the default version of the TALYS 1.96 nuclear reaction code, the maximum number of discrete levels that are considered in the Hauser-Feshbach model for residual nuclei is set to 30. High-Z nuclei such as Fe and Cu, which exhibit non-negligible neutron yields from natural radioactivity, typically have more than 30 excited states that contribute to neutron production via interactions with high-energy alpha particles. Increasing the number of excited states used in the library allows more accurate calculation of neutron energy spectra without affecting the total neutron yield. We have modified the TALYS code in order to increase the number of levels to include all the possible excited states for the final-state nucleus that are available in RIPL-3 \cite{RIPL-3} library.

In this work we report the very latest amendments to the libraries of SOURCES4A that include cross-sections calculated with the latest version of TALYS 1.96~\cite{Koning:2023ixl} with the extended range of excited states and evaluated data from JENDL-5 \cite{JENDL5}, as well as some very recent experimental data. We have also optimised the choice of cross-sections for neutron yield calculations to get a better agreement with experimental data including those from alpha beams and radioactive decay chains.

Section~\ref{Cross-sections-comp} presents a comparison of our 'optimised' cross-sections with measurements and results from the statistical codes TALYS and EMPIRE, along with data from the JENDL database, as a few examples relevant to underground experiments aimed at detecting rare events.

\section{Cross sections comparison}
\label{Cross-sections-comp}

The cross-sections from different models and experimental data were compared in Ref.~\cite{Kudryavtsev:2020eer}. The same paper included calculations of neutron yields with SOURCES4A using different models for the cross-sections. Here we extend previous work and present neutron yield calculations with optimised cross-sections as explained in 
Section~\ref{sec:sources4a} comparing this to available experimental data. 
We focus on materials extensively used in rare event physics experiments, such as those searching for dark matter, neutrino interactions, or other phenomena beyond the standard model.

The list of other elements with updated cross-sections in this study comprises Be, B, N, Na, Ca, P, Cl, Mg, Ti, Fe, Ni, Cr, Mn, Cu, V, Mo.

\subsection{Lithium}
Recently, the CRESST Collaboration suggested the idea of using LiAlO$_2$ crystals for direct dark matter searches due to the fact that $^7$Li is a good candidate for studying spin-dependent dark matter interactions in the sub-GeV mass region. $^6$Li also has a high cross-section for neutron capture providing a unique signature that allows monitoring of the neutron flux on site \cite{bertoldo2020lithium,CRESST:2020tlq}.

\begin{figure}[ht]
  \centering
  \begin{subfigure}{.5\textwidth}
    \includegraphics[width=1.\linewidth]{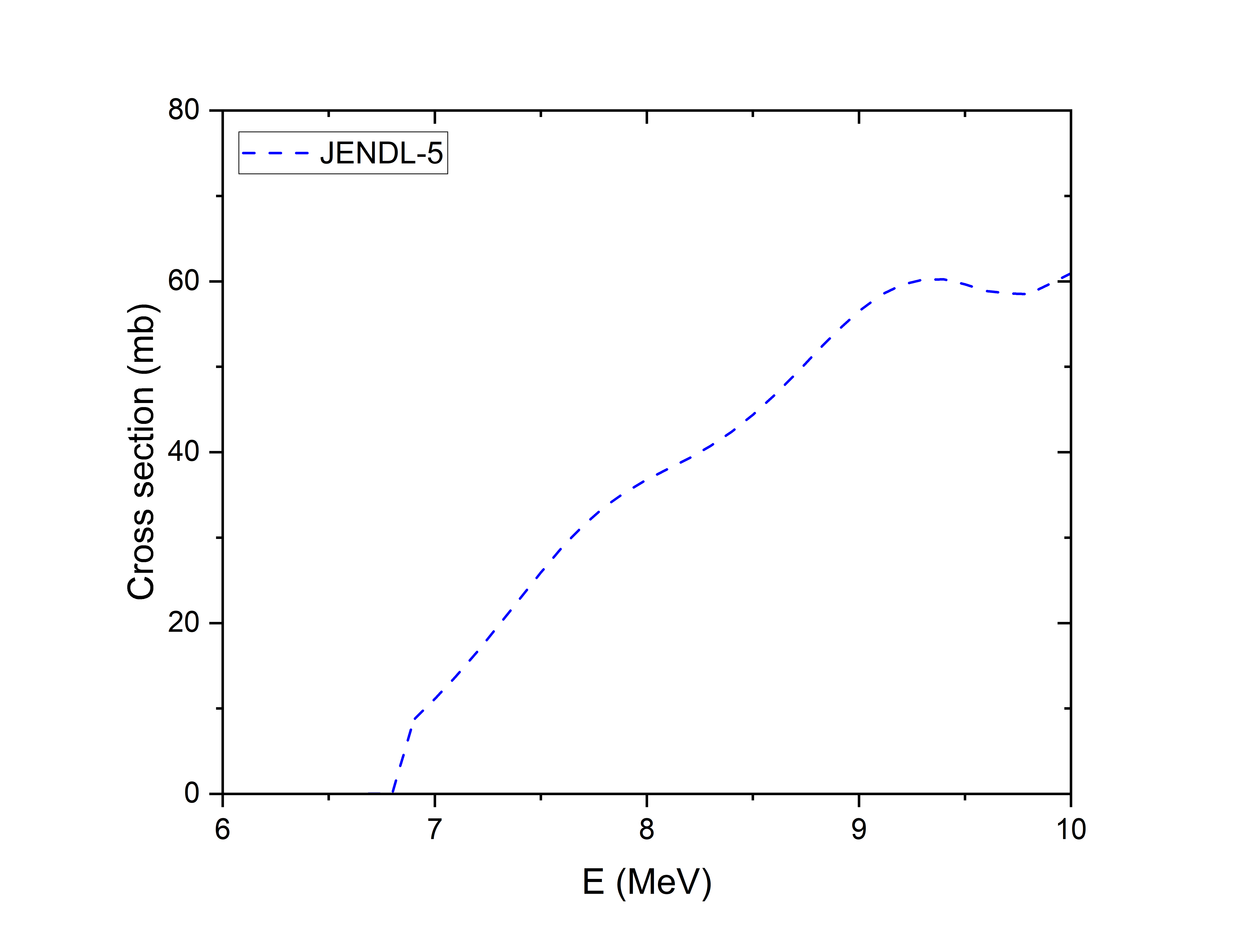}
    \caption{ }
    \label{fig:xs-6Li}
  \end{subfigure}%
  \begin{subfigure}{.5\textwidth}
    \includegraphics[width=1.\linewidth]{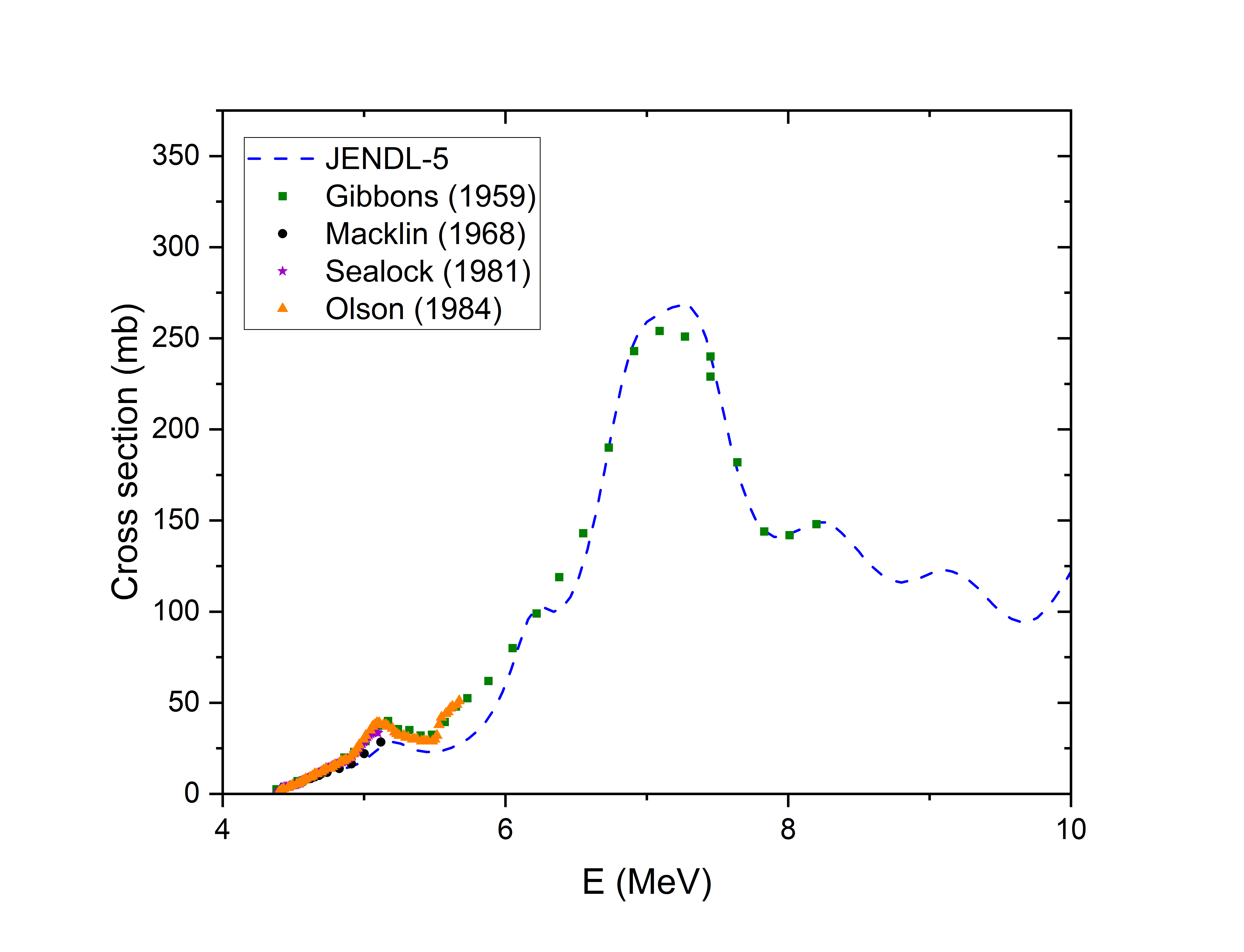}
    \caption{ }
    \label{fig:xs-7Li}
  \end{subfigure}
  \caption{(a) $^{6}$Li($\alpha$,n)$^{9}$B cross section as a function of alpha energy. 
  %In this case, there is no experimental data available. 
  (b) $^{7}$Li($\alpha$,n)$^{10}$B cross section as a function of alpha energy. Experimental data for $^7$Li are taken from \cite{PhysRev.114.571,PhysRev.165.1147,SEALOCK1981279,PhysRevC.30.1375}. Uncertainties of the measurements are not shown on this and other figures to make data points and calculations clearly visible; these uncertainties can be found in the original publications.}
\label{Li7}
\end{figure}

Excitation functions for $^6$Li($\alpha,n$)$^9$B and $^7$Li($\alpha,n$)$^{10}$B reactions from threshold to 15 MeV were reported in Ref.~\cite{MEHTA196390} and included in JENDL-5 library. Total cross-sections were evaluated to match the measurements of thick target neutron yields from Ref.~\cite{bair1979neutron}. These cross-sections are shown in Figure \ref{Li7}) together with other data available for $^7$Li.

\subsection{Carbon}
Carbon contains 1.07\% of $^{13}$C and is present in scintillators, various %composition of 
plastics and other polymers, such as acrylic, polyethylene, nylon and PTFE. These materials are often placed near the target volume in rare event experiments. In addition, carbon is prevalent in the composition of rocks.

\begin{figure}[ht]
\centering
\includegraphics[width=.8\linewidth]{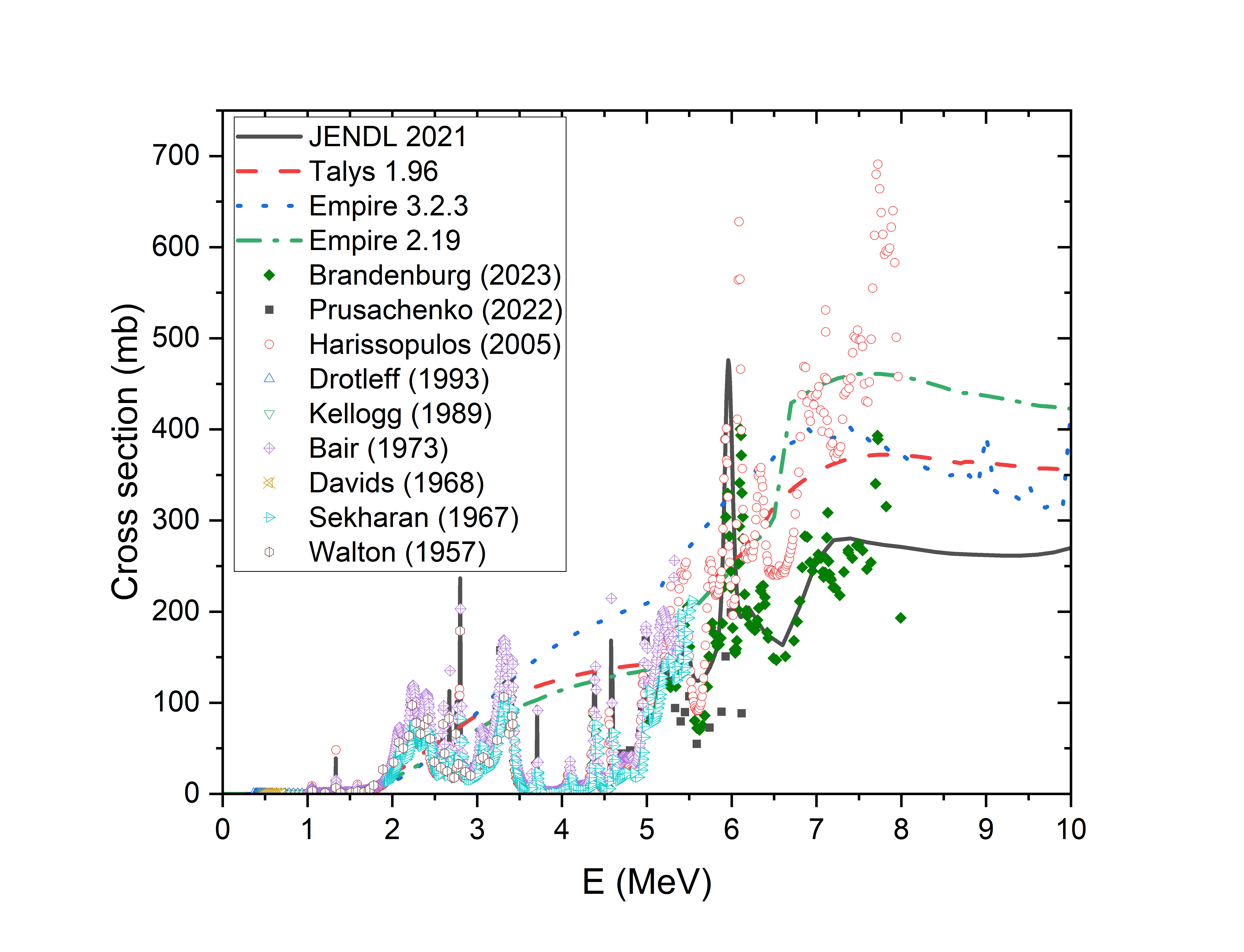}
\caption{$^{13}$C($\alpha$,n)$^{16}$O cross section as a function of alpha energy. Experimental data are taken from \cite{PhysRevC.108.L061601,Prusachenko2005,harissopulos2005,drotleff1993,kellogg1989,Bair_Haas_1973,DAVIDS1968619,shekharan1967,Walton1957}. $^{12}$C does not contribute to neutron yield since the energy threshold of the corresponding reaction exceeds the maximum energy of alphas from radioactive decay chains of natural uranium and thorium.}
\label{fig:xs-13C}
\end{figure}

%The most recent experimental data available for the cross sections from %Prusachenko et al.
%Ref.~\cite{Prusachenko2005} are only for the ground state population. Thus, for this isotope, 
The most recent measurements from Ref.~\cite{PhysRevC.108.L061601} are used in this work from 2.88~MeV up to 8~MeV and complemented with the cross-sections from~\cite{harissopulos2005} below 2.88~MeV and from JENDL library above 8 MeV
%TALYS~1.96 calculations above this energy 
(see Figure~\ref{fig:xs-13C}). Experimental uncertainties are not shown on the figures and can be found in the original publications. As an example, the uncertainties for $^{13}$C~\cite{10.1117/12.204163} are dominated by systematic uncertainties of about 16\%.

\subsection{Oxygen}
%Oxygen is another element present in experimental setups. 
Oxygen is present in water, water-based liquid scintillators (WbLS) \cite{wbls_sym15010011}, some plastics, and rocks. 
Two isotopes are of interest for neutron production in ($\alpha,n$) reactions: $^{17}$O and $^{18}$O with abundances of 0.038\% and 0.205\%, respectively. $^{16}$O has an energy threshold for ($\alpha,n$) reaction above the energies of alphas from radioactivity.

\begin{figure}[!ht]
  \centering
  \begin{subfigure}{.5\textwidth}
    \includegraphics[width=1.\linewidth]{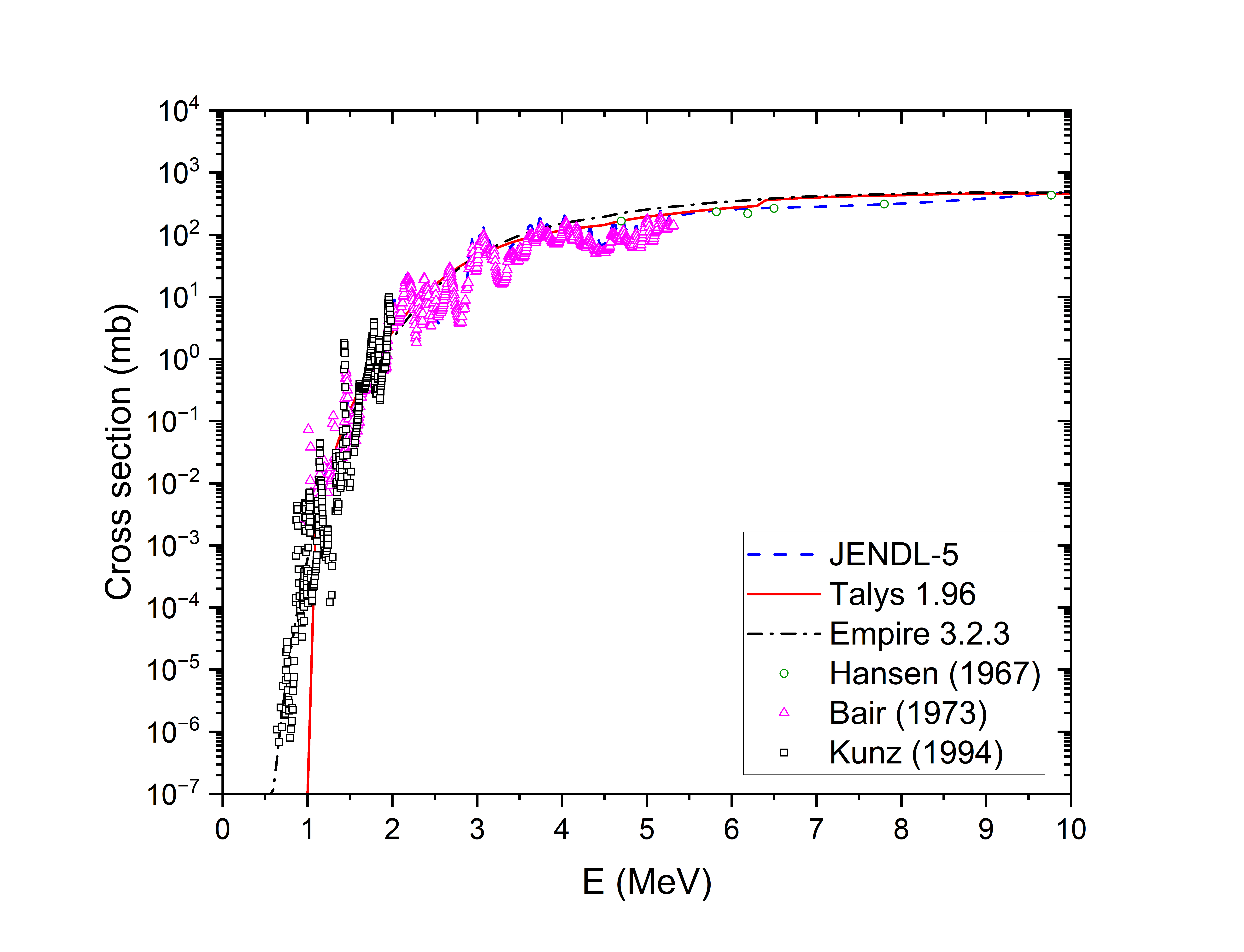}
    \caption{ }
    \label{fig:xs-17O}
  \end{subfigure}%
  \begin{subfigure}{.5\textwidth}
    \includegraphics[width=1.\linewidth]{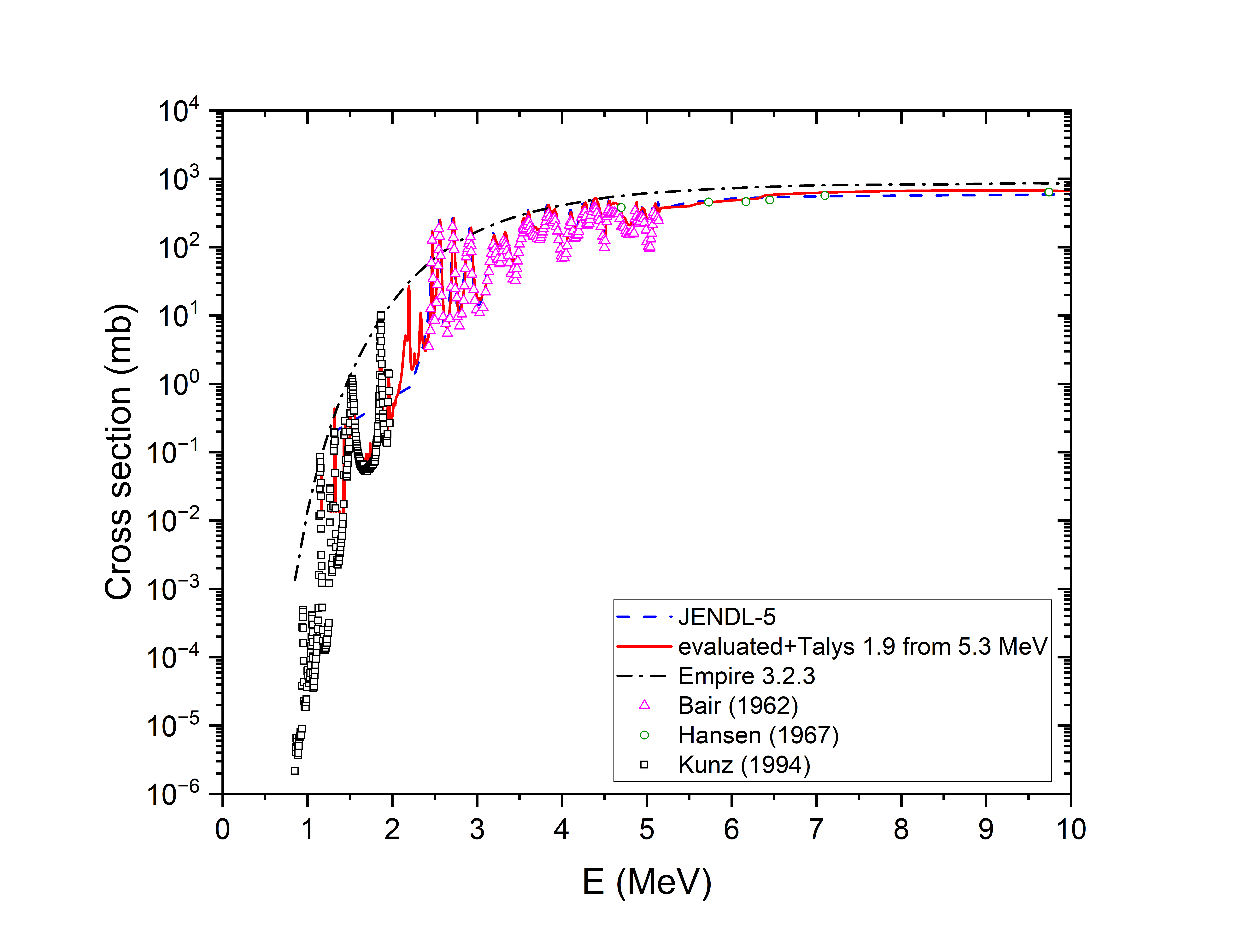}
    \caption{ }
    \label{fig:xs-18O}
  \end{subfigure}
  \caption{(a) $^{17}$O($\alpha$,n)$^{20}$Ne cross section as a function of alpha energy. (b) $^{18}$O($\alpha$,n)$^{21}$Ne cross section as a function of alpha energy. Experimental data are taken from \cite{HANSEN196725,Bair_Haas_1973,10.1117/12.204163}.}
\end{figure}

For $^{17}$O, the measurements from Ref.~\cite{10.1117/12.204163,Bair_Haas_1973} have been used up to 5.2~MeV and complemented with the JENDL-5 data above this energy (see Figure~\ref{fig:xs-17O}). In the case of $^{18}$O, the cross sections evaluated using JENDL-5 have been used (see Figure \ref{fig:xs-18O}).

\subsection{Fluorine}
\begin{figure}[!ht]
  \centering
    \includegraphics[width=.8\linewidth]{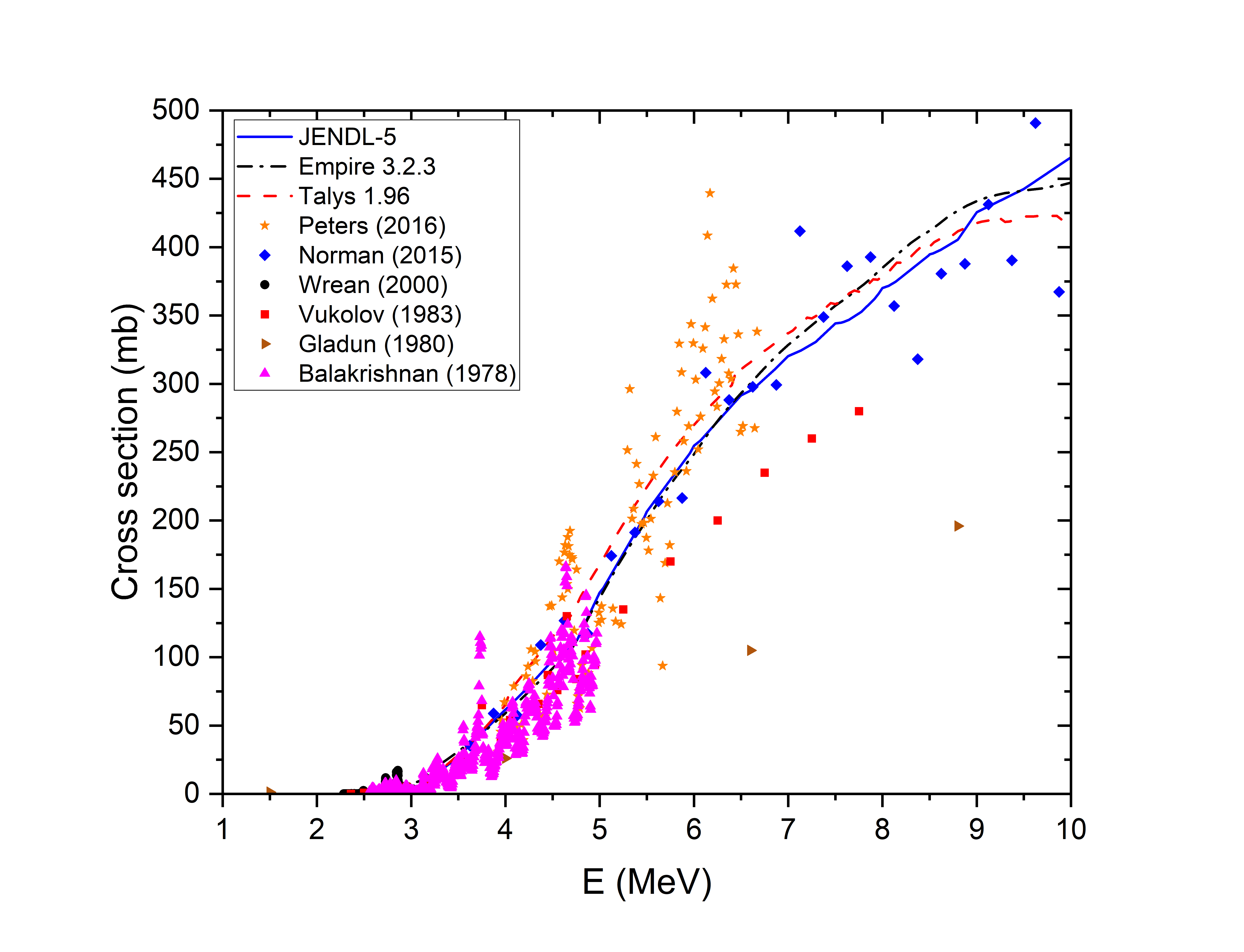}
  \caption{(a) $^{19}$F($\alpha$,n)$^{22}$Na cross section as a function of alpha energy. Experimental data are taken from \cite{peters2016,norman2015,wrean2000,vukolov1983,gladun1980,balakrishnan1978}.}
\label{fig:xs-19F}
\end{figure}

Polytetrafluoroethylene (PTFE) is widely used %as a reflector 
in rare event 
xenon or argon experiments 
%using xenon or argon as active medium 
because it withstands temperatures of liquid xenon (LXe) and liquid argon (LAr) and is a good insulator \cite{aprile2011material,AGNES2015456,XENON:2021mrg}. It is also an excellent reflector for VUV light at the LXe and LAr scintillation wavelengths. In this case, the measurements %of Peters 
from Ref.~\cite{peters2016} were used up to 7 MeV and complemented with TALYS 1.96 calculations above this energy. Figure \ref{fig:xs-19F} shows a comparison of the experimental cross-sections with calculations.

\subsection{Aluminium}

%Aluminium is used in cryostats, shielding, support structures, and purification systems. 
Sapphire crystals containing aluminium have been proposed for light dark matter detection 
%by scattering off electrons 
\cite{hochberg2016detecting,hochbergPhysRevLett}.
Aluminium is also used in the readout electrodes and other components.

\begin{figure}[!ht]
  \centering
    \includegraphics[width=.8\linewidth]{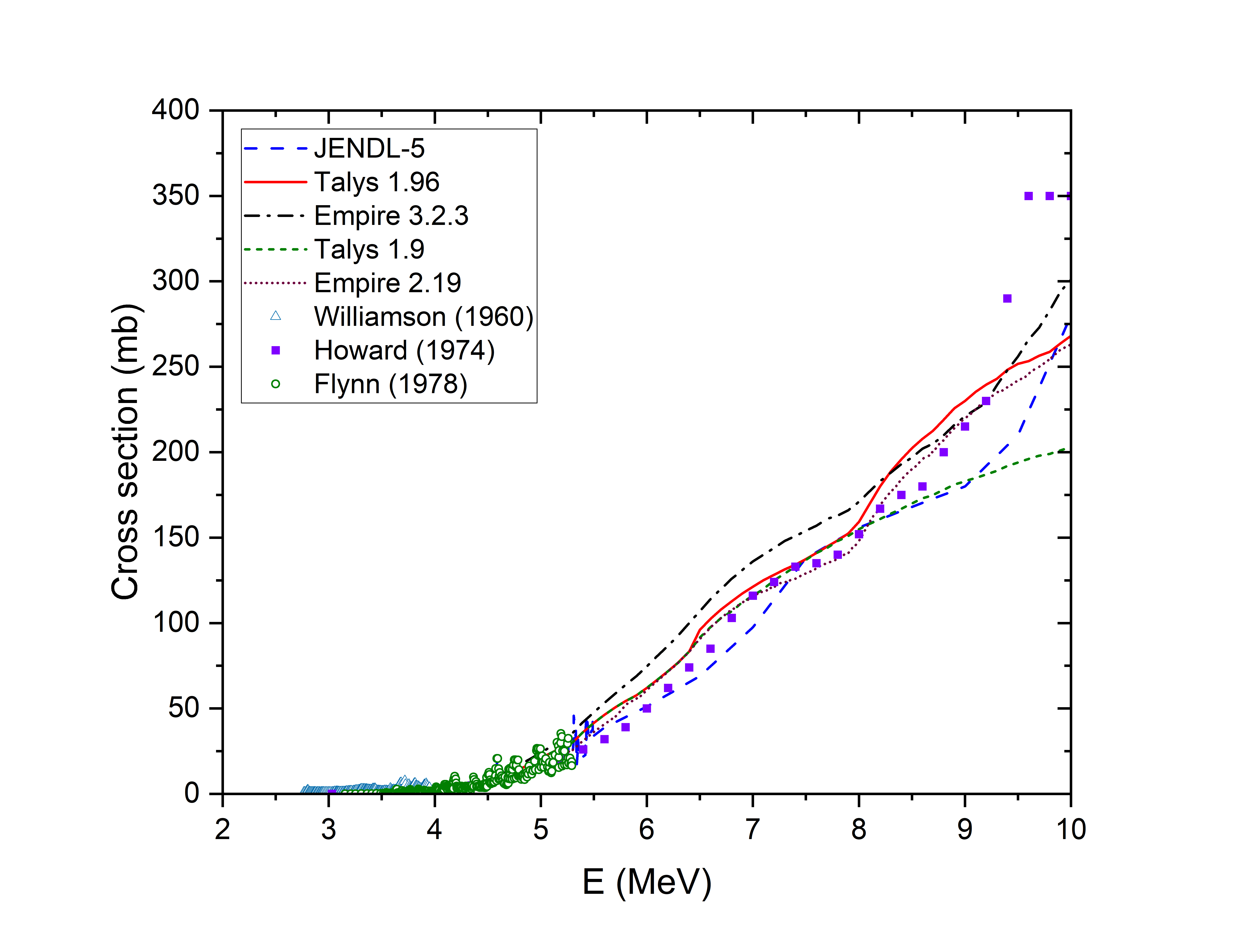}
  \caption{$^{27}$Al($\alpha$,n)$^{30}$P cross section as a function of alpha energy. Experimental data are taken from \cite{williamson1960,howard1974,flynn1978}.}
\label{fig:xs-27Al}
\end{figure}

For aluminium, the cross-sections measured in Refs.~\cite{flynn1978,howard1974} were used and complemented with those calculated with EMPIRE~3.2.3 above 9.2~MeV (see Figure \ref{fig:xs-27Al}). This energy is already above the highest alpha energy from radioactive decay chains. Experimental data at high energies may be contaminated by (alpha,pn) reaction \cite{PhysRev.133.B911}.

\subsection{Silicon}

Silicon crystals of high purity are employed as substrates for detectors in dark matter direct detection experiments, such as CDMS \cite{SHUTT1996318}, CDMS-II \cite{Brink_2009} and SuperCDMS \cite{Agnese_2017}.
Silicon is also present in quartz, glass and light sensors.

\begin{figure}[!ht]
  \centering
    \includegraphics[width=.8\linewidth]{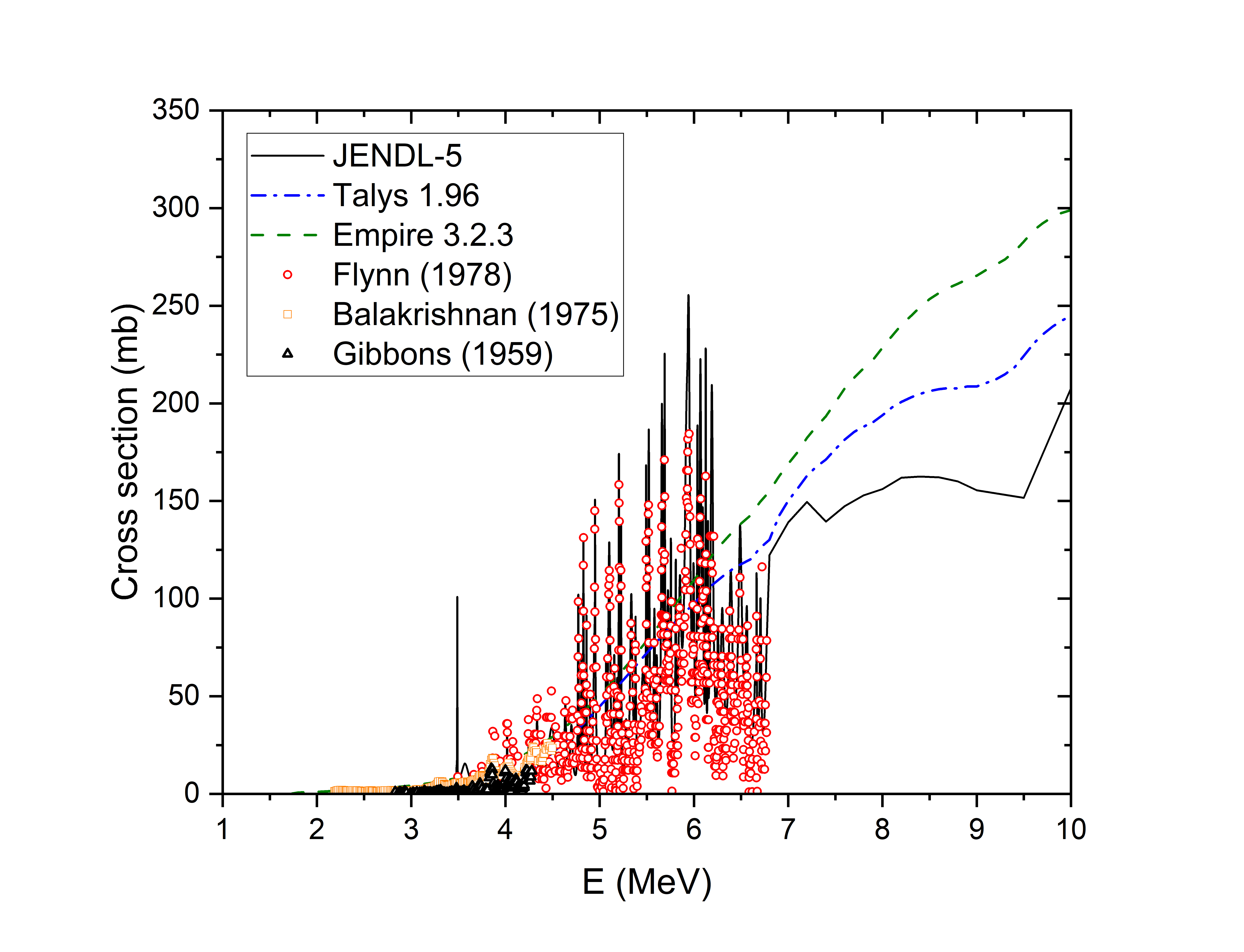}
    \caption{$^{29}$Si($\alpha$,n)$^{32}$S cross section as a function of alpha energy. Experimental data are taken from Refs.~\cite{flynn1978,PhysRevC.11.54,PhysRev.114.571}.}
    \label{fig:xs-29Si}
\end{figure}

For $^{29}$Si, the measured cross-section from Ref.~\cite{flynn1978} was used up to 6.8~MeV and the TALYS 1.96 cross-section above this energy (see Figure \ref{fig:xs-29Si}). As discussed above, none of the statistical models used in the codes can accurately predict the resonance shape of the cross-section and the use of experimental data is unavoidable. Experimental data sets meanwhile are consistent and can be used up to energies where they are available (6.8~MeV). Above this energy, the choice of the cross-section is quite arbitrary and an option of using TALYS 1.96 was dictated by: a) the most recent update to the calculations and b) a reasonable agreement with experimental data at energies close to the upper end of data range.

\subsection{Noble elements used in neutrino and dark matter experiments}

Among the noble elements, argon and xenon are the only ones currently used as targets for neutrino and dark matter 
experiments
%detection 
due to their high scintillation yield, no (xenon) or relatively low intrinsic radioactivity (argon), and excellent electron transport characteristics \cite{BAUDIS201450}. Neon has been proposed for detecting low-energy neutrinos and might also be used for WIMPs searches \cite{PhysRevC.86.015807}. 

For all isotopes of xenon, the energy threshold for ($\alpha,n$) reactions, which is determined by the Coulomb barrier for high-$Z$ materials, is above 9 MeV. 
These energies are higher than those of $\alpha$-particles from radioactive decay chains.

Argon is widely used as active volume in rare event searches and several examples include DUNE \cite{Abi_2020}, DarkSide \cite{Aalseth_2020} and ArDM \cite{Regenfus_2010} %and MicroBooNE \cite{Acciarri_2017}. 
For $\alpha$-particle energies up to 10 MeV, the energy dependence of the ($\alpha,n$) reaction cross-sections calculated with TALYS and EMPIRE nuclear reaction codes agree reasonably well. However, there is a significant discrepancy between the calculated values and the experimental data point. This measurement has been done at 90 degree neutron scattering angle with respect to the alpha beam and extrapolated to the total cross-section assuming isotropic neutron emission. The authors of the measurement estimated the uncertainty to be up to a factor of 2 but the difference with the models suggest that the uncertainty, in particular due to the assumption of isotropic neutron emission, may be much larger. In this case, we used EMPIRE 2.19 because its calculation is closer to the only available experimental data point.

For the materials discussed above, as well others considered in this study, the choice of optimum cross-section was dictated by the consistency between different data sets, agreement with models where model predictions are reliable and agreement between neutron yield calculations with alpha beam and radioactivity data with the last topic being discussed in the next Section. If no choice can be made based on this approach, we selected the most recent data or calculations, or the highest cross-section to be conservative in estimating the neutron yield bearing in mind the main goal of estimating background for rare event searches.

\begin{figure}[!ht]
  \centering
    \includegraphics[width=.8\linewidth]{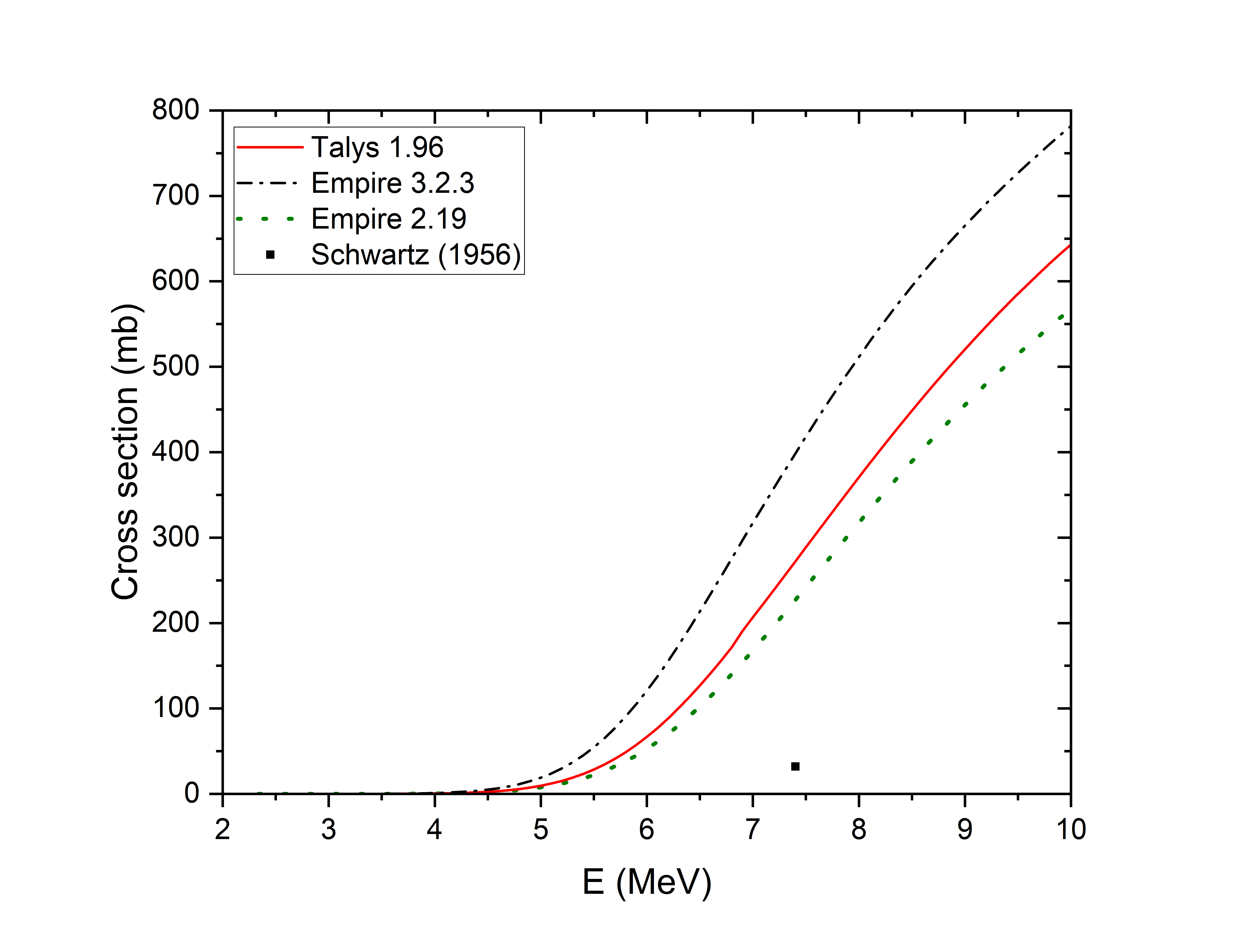}
  \caption{$^{40}$Ar($\alpha$,n)$^{43}$Ca cross-section as a function of alpha energy. Experimental data is taken from Ref.~\cite{PhysRev.101.1370}.}
\label{fig:xs-40Al}
\end{figure}

\section{Comparison of SOURCES4 with experimental data}
\label{sec:comparison}

\begin{figure}[ht]
  \centering
    \includegraphics[width=.8\linewidth]{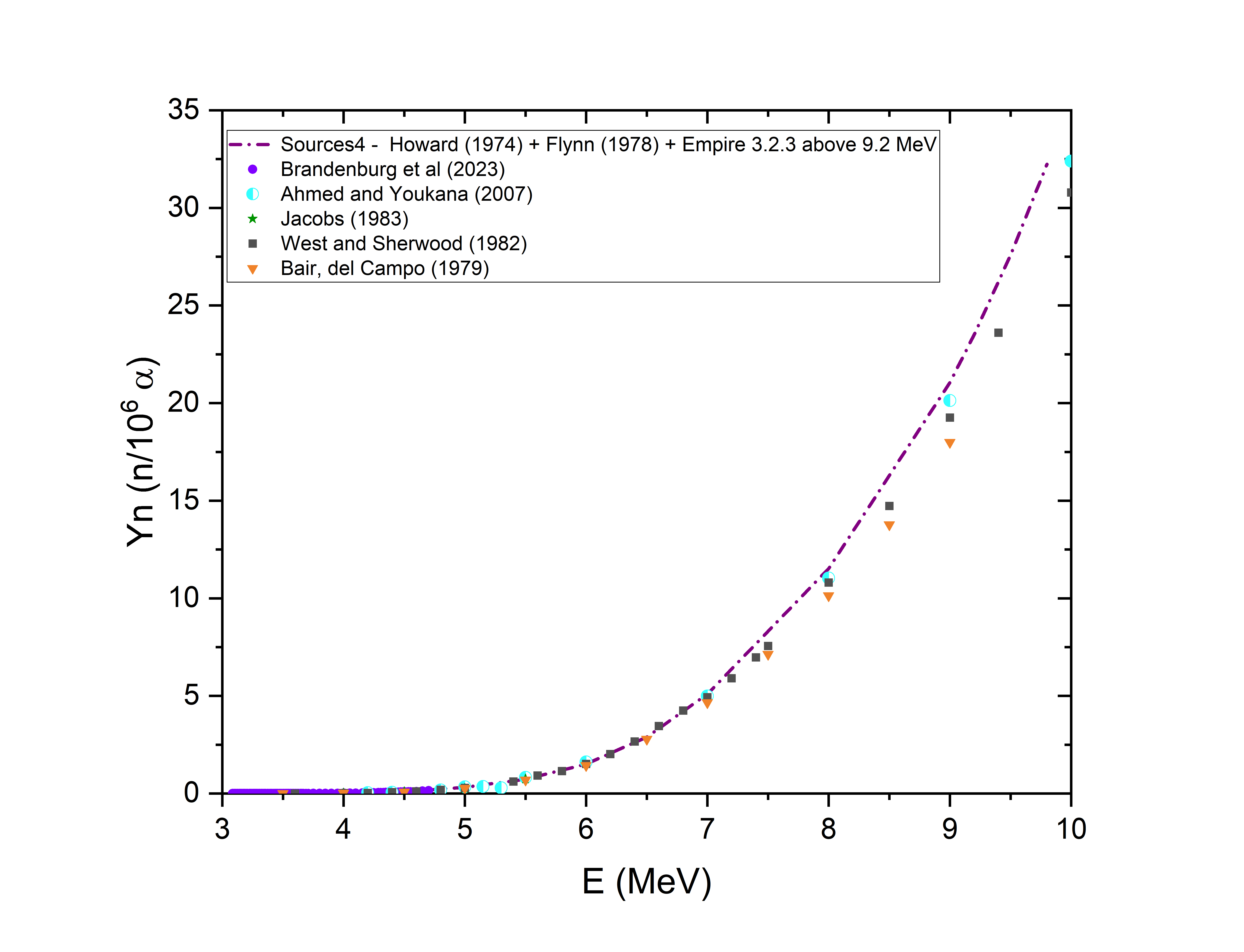}
  \caption{$^{27}$Al($\alpha$,n)$^{30}$P neutron yield as a function of alpha energy. Optimised cross-sections are used in SOURCES4A. Experimental data are taken from \cite{Brandenburg_2022,AHMED2007287,JACOBS1983541,west1982,bair1979neutron}.}
 \label{fig:Yn-27Al}
\end{figure}

\begin{figure}[!ht]
  \centering
    \includegraphics[width=.8\linewidth]{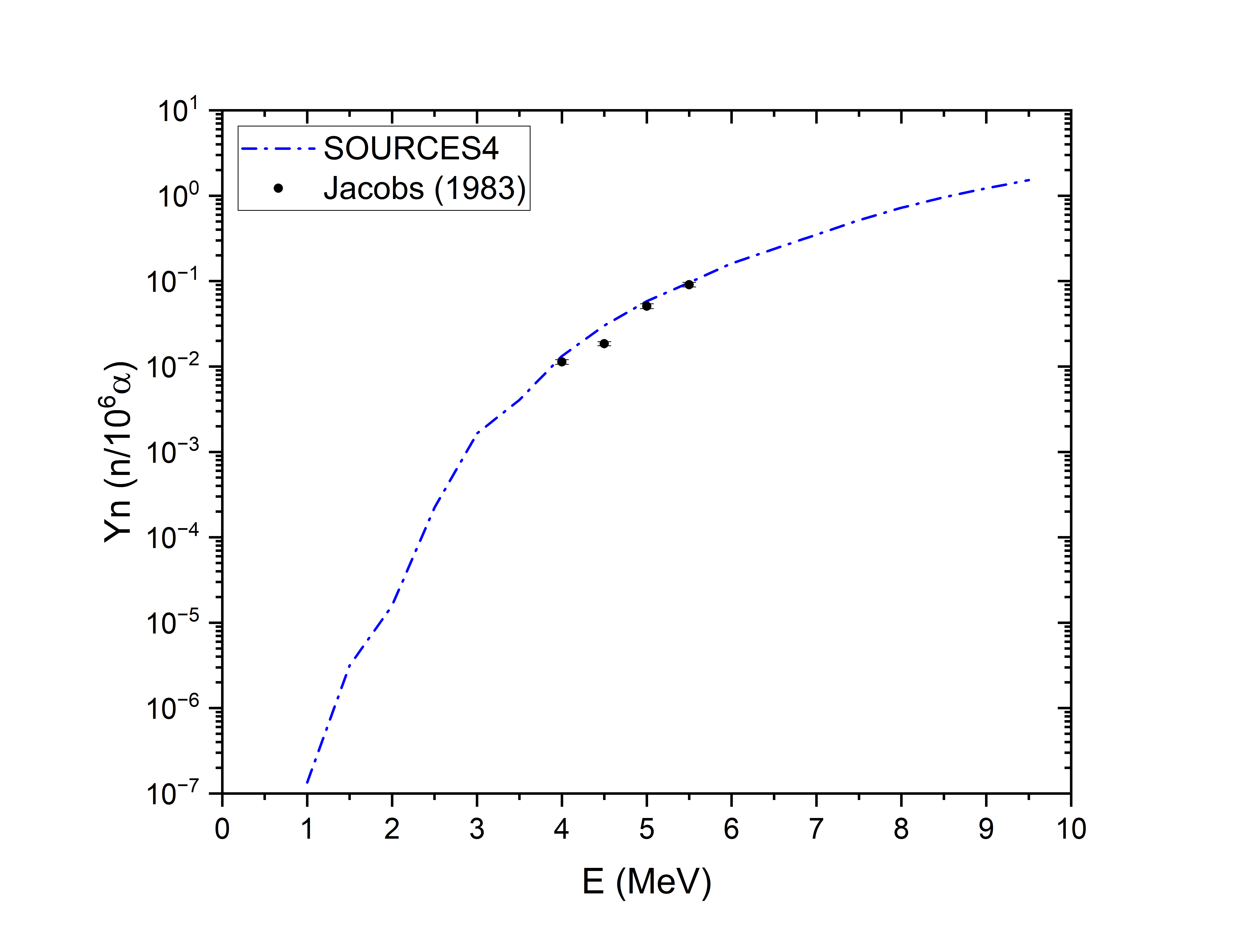}
    \caption{SiO$_2$ neutron yield as a function of alpha energy. Optimised cross-sections are used in SOURCES4A. Experimental data are taken from \cite{JACOBS1983541}.}
    \label{fig:Yn-SiO2_logy}
\end{figure}

Figures~\ref{fig:Yn-27Al} and \ref{fig:Yn-SiO2_logy} show the dependence of the neutron yield on the alpha energy for two materials, as examples, that are frequently used in underground experiments, aluminium and silicon oxide. Following a similar approach as in previous publications \cite{fernandes,mendoza2019}, the neutron yield is given as the number of neutrons per $10^6$ alphas. In our calculations with SOURCES4 we use optimised cross-sections - a combination of experimentally measured cross-sections and calculations or evaluations for energies where data are not available, as described in Section \ref{Cross-sections-comp}. A good agreement with measurements is observed for all materials tested confirming the right choice of cross-sections for the isotopes in these materials. Comparing the results from SOURCES4 calculations for $^{27}$Al for different alpha energies with available experimental data \cite{bair1979neutron,west1982,AHMED2007287}, we found that the ratio of calculated values to the measurements varies from 0.93 to 1.17 showing a good agreement between them across a large range of alpha energies.
%given typical experimental uncertainties of ~20\%.

\begin{figure}[!ht]
  \centering
  \begin{subfigure}{.5\textwidth}
    \includegraphics[width=1.\linewidth]{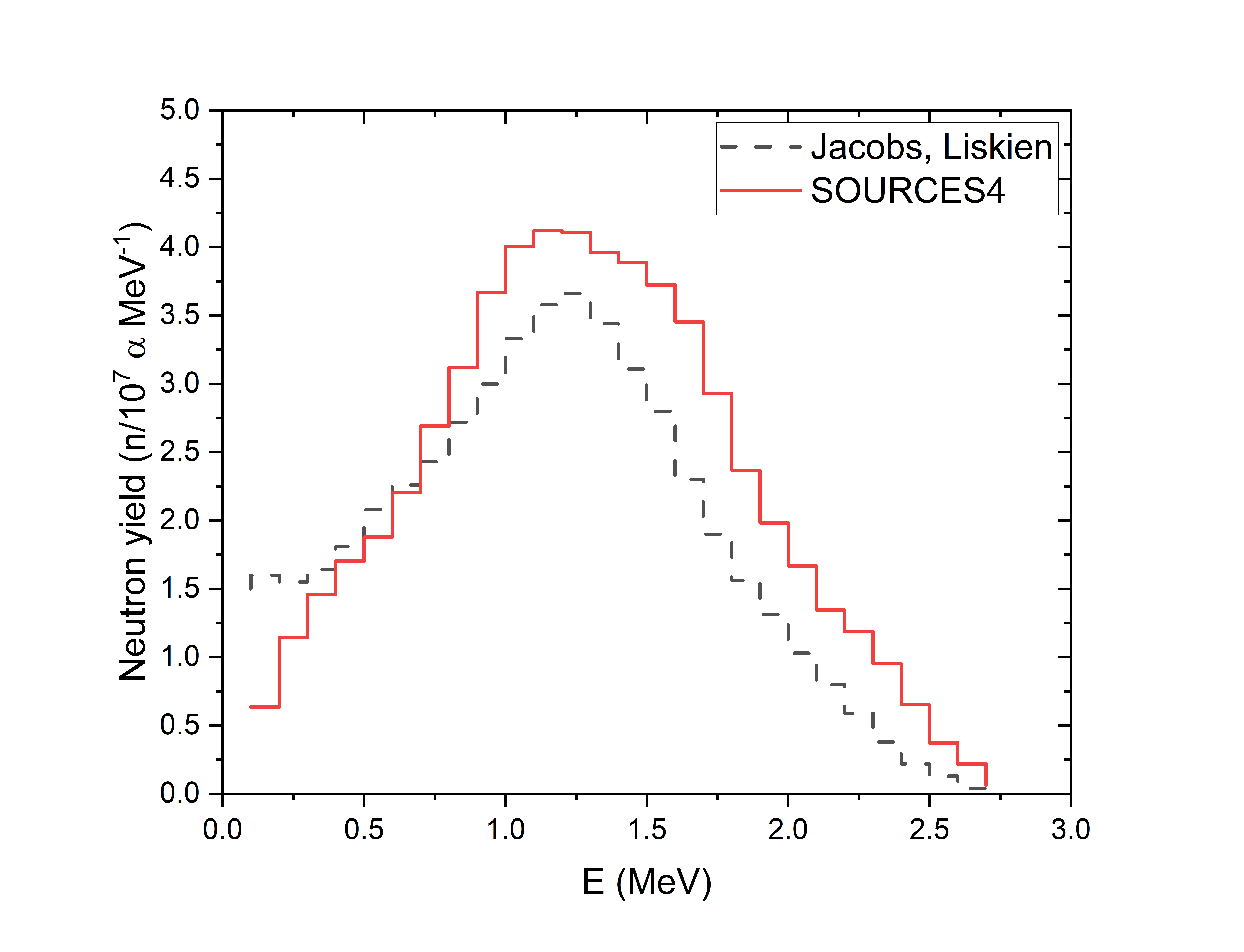}
    \caption{ }
    \label{fig:n_spectr_19F}
  \end{subfigure}%
  \begin{subfigure}{.5\textwidth}
    \includegraphics[width=1.\linewidth]{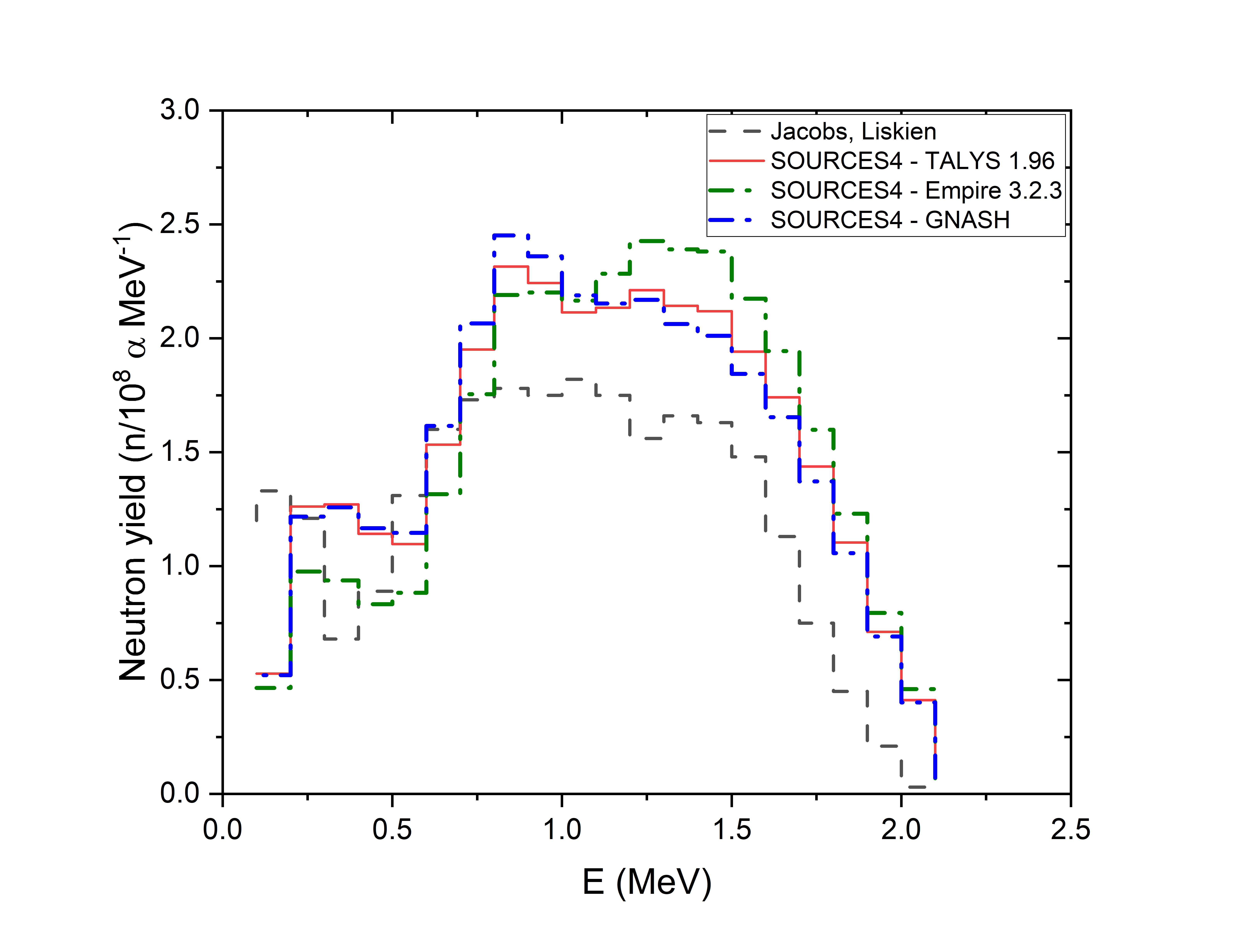}
    \caption{  }
    \label{fig:n_spectr_27Al}
  \end{subfigure}
  \caption{Neutron energy spectrum from 5 MeV $\alpha$ particles in (a) $^{19}$F and (b) $^{27}$Al in comparison with data from Ref. \cite{JACOBS1983541}. Branching ratios from TALYS 1.96 has been used for $^{19}$F and three different models have been compared for $^{27}$Al.}
  \label{19F}
\end{figure}

Neutron energy spectra as calculated with SOURCES4 and measured in Ref.~\cite{JACOBS1983541} for 5~MeV $\alpha$ particles incident on fluorine and aluminium targets are shown in Figure~\ref{19F}. The shape of the spectra are reproduced correctly in the calculations and, given the difficulty to accurately reconstruct neutron energy in experiments the overall agreement is reasonably good. Also, there are very few measurements of the branching ratios for transitions to the ground and excited states and only for low Z targets, so we are almost fully relying here on models. Given that statistical models do not predict correctly the resonant structure of the cross-section, they are not expected to give an accurate calculation for branching ratios either.  Hence, within possible uncertainties, the agreement looks reasonably good.

\begin{figure}[!ht]
  \centering
  \begin{subfigure}{.5\textwidth}
    \includegraphics[width=1.\linewidth]{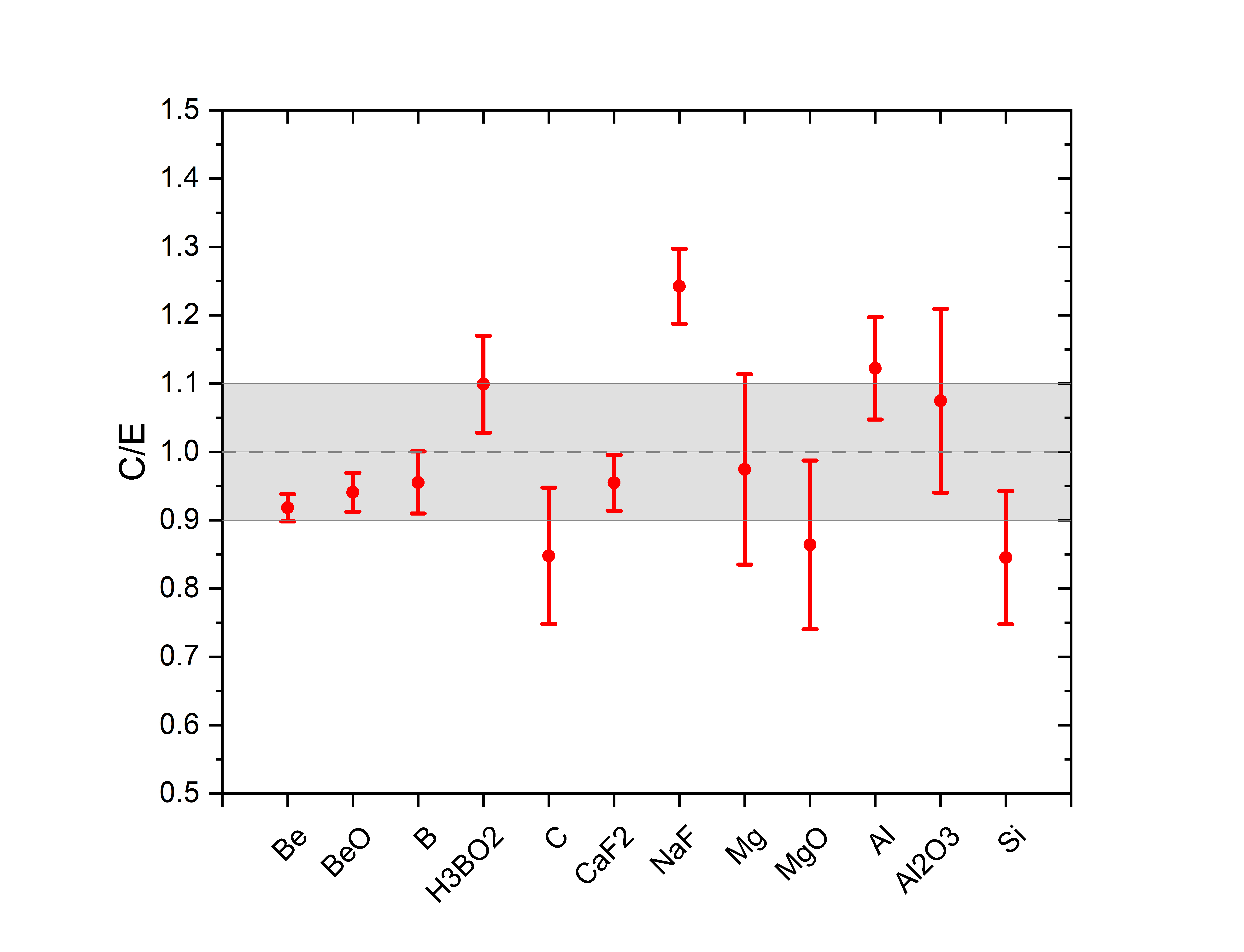}
    \caption{ }
    \label{fig:ratio-U-Gorshkov}
  \end{subfigure}%
  \begin{subfigure}{.5\textwidth}
    \includegraphics[width=1.\linewidth]{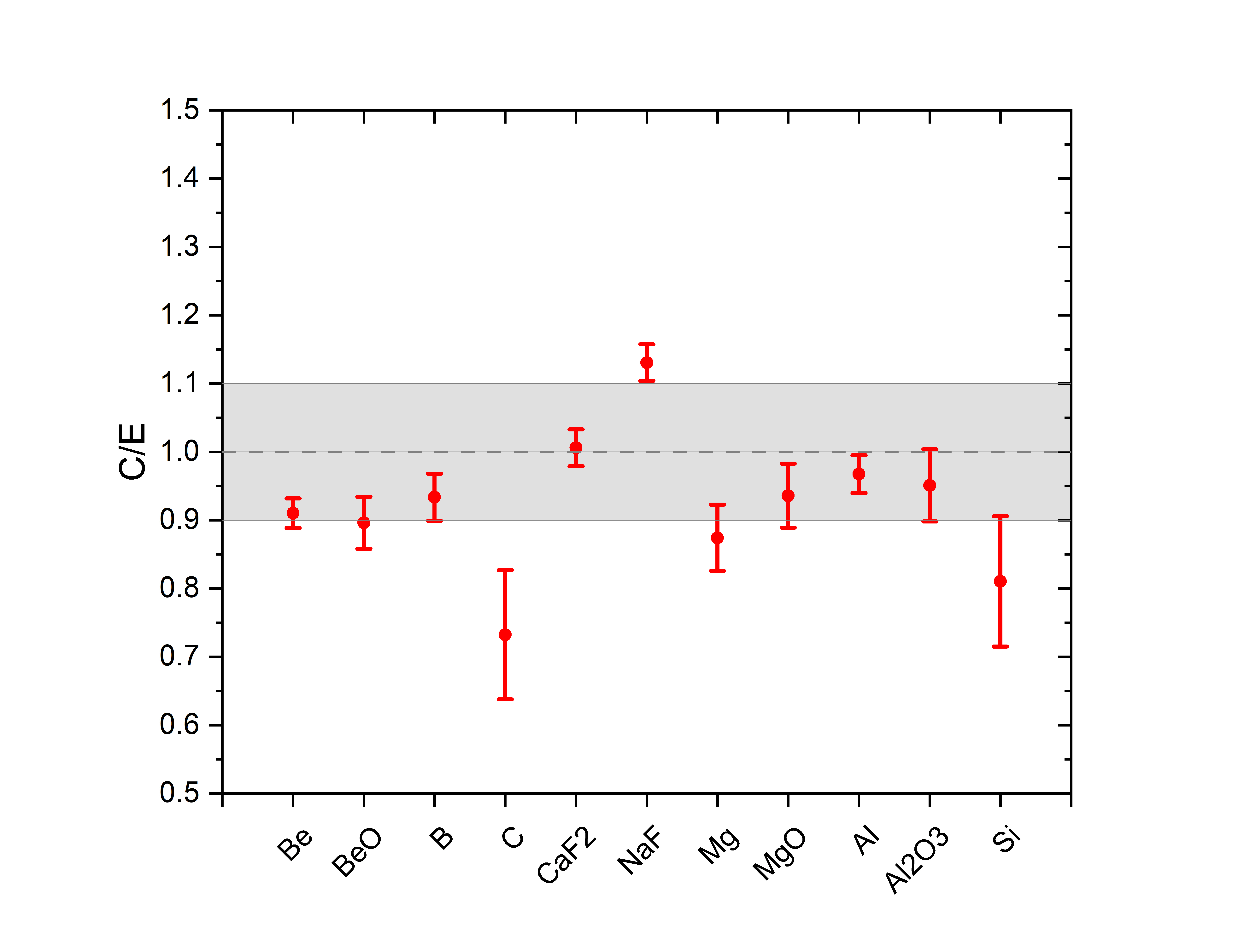}
    \caption{ }
    \label{fig:ratio-Th-Gorshkov}
  \end{subfigure}
  \caption{Ratio of the calculations (C) with optimised SOURCES4 to the neutron yield measured in Ref.~\cite{gorshkov1964b} (E) for: (a) $^{\mathrm{nat}}$U and (b) $^{232}$Th decay chains in several materials. The shaded area represents the $\pm 10\%$ difference. Only the experimental errors are included.}
  \label{fig:Gorshkov}
\end{figure}

\begin{figure}[!ht]
  \centering
  \begin{subfigure}{.5\textwidth}
    \includegraphics[width=1.\linewidth]{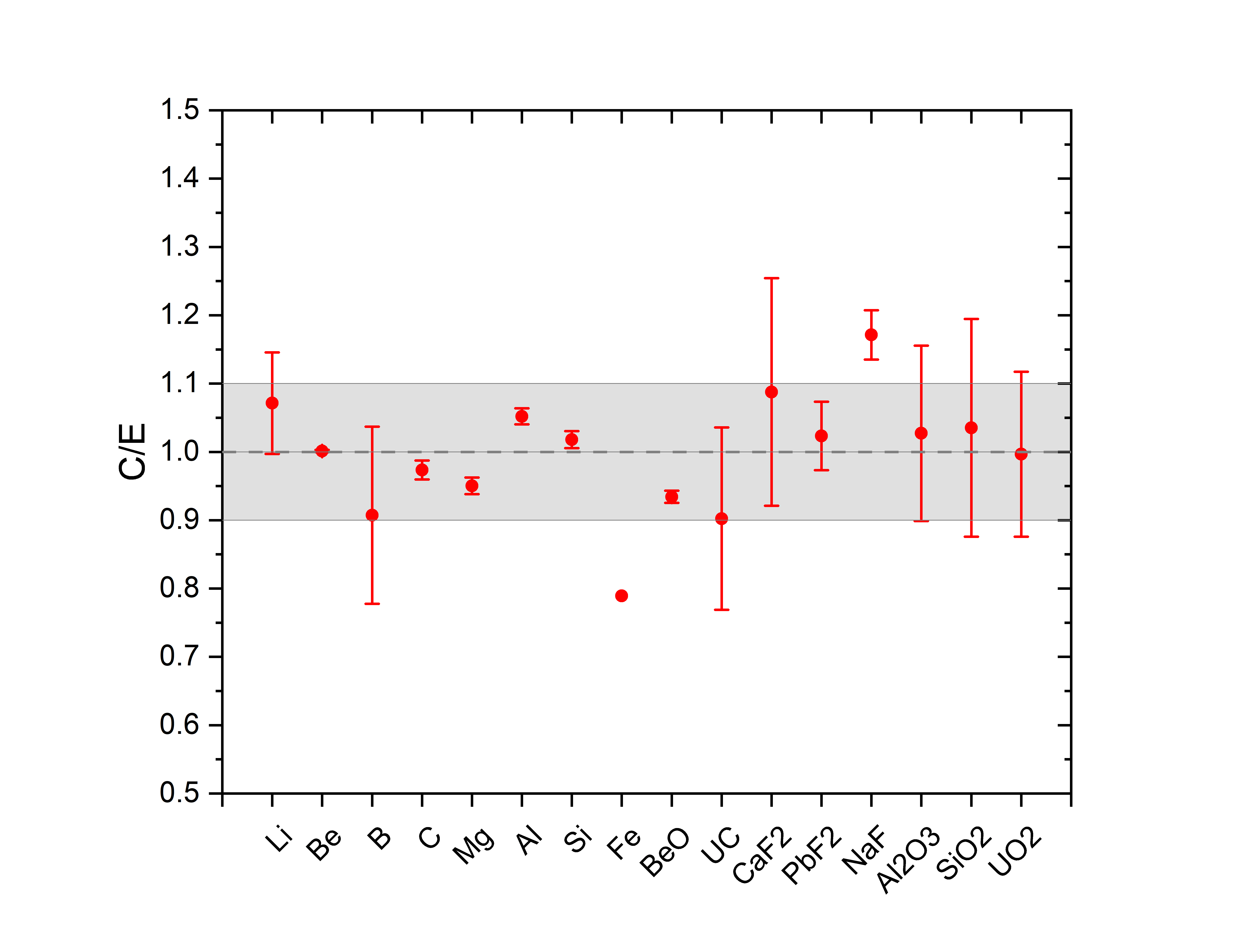}
    \caption{ }
    \label{fig:238U_Fernandez}
  \end{subfigure}%
  \begin{subfigure}{.5\textwidth}
    \includegraphics[width=1.\linewidth]{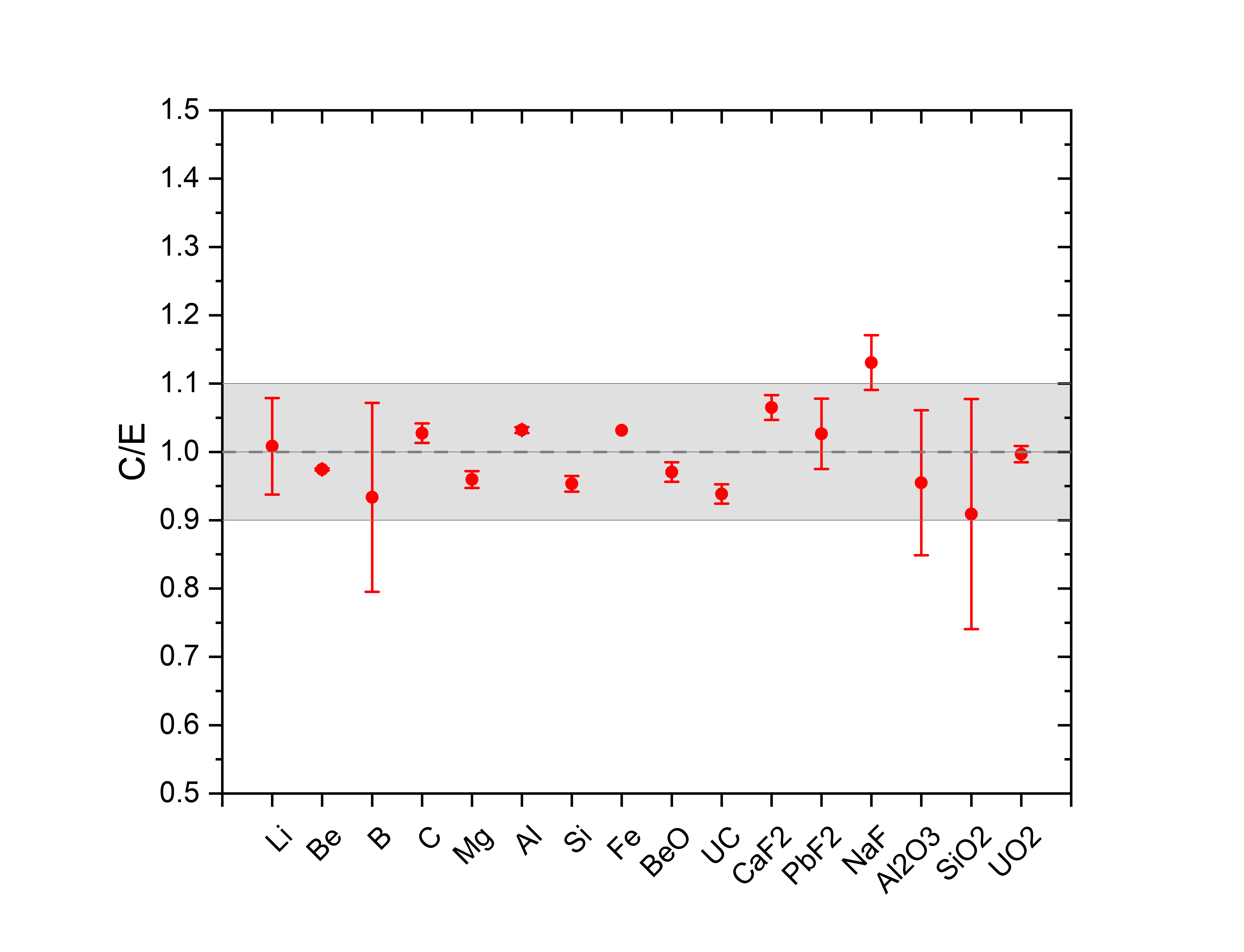}
    \caption{ }
    \label{fig:232Th_Fernandez}
  \end{subfigure}
  \caption{Ratio of the calculated values (C) from the optimised SOURCES4 code to the evaluated neutron yield data (E) from Ref.~\cite{fernandes} for (a) $^{238}$U and (b) $^{232}$Th decay chains in several materials. The shaded area represents the $\pm 10\%$ difference. No error was given for Fe. Only the experimental errors were taken into account for other materials.}
  \label{fig:Fernandez}
\end{figure}

In underground experiments the prime interest is in the evaluation of neutron yields and energy spectra in ($\alpha,n$) reactions from naturally occurring radioactive isotopes and decay chains. 
A comparison of the calculations with measurements from Ref.~\cite{gorshkov1964b} for the whole uranium and thorium decay chains is shown in Figure~\ref{fig:Gorshkov} whereas a similar comparison with evaluated data from Ref.~\cite{fernandes} is included in Figure \ref{fig:Fernandez}. Only experimental uncertainties are shown as quoted in Refs.~\cite{gorshkov1964b,fernandes}.

For most materials the agreement between calculations and data is within 10\%, again confirming correct choice of cross-sections in SOURCES4 library. 
Neutron yields for carbon agree with the measurements reported in Ref.~\cite{fernandes} (Figure~\ref{fig:Fernandez}) but are lower than those reported in Ref.~\cite{gorshkov1964b}. We note that the data in Ref.~\cite{fernandes} are not the direct measurements of neutron yields from uranium and thorium decay chains but were extrapolated from neutron yields measured with $\alpha$ particle beams taking into account the $\alpha$ energies and energy losses in different materials leading to an additional uncertainty in the results. In our calculations we have used the most recent measurement of the cross-section for $^{13}$C \cite{PhysRevC.108.L061601} which is lower than the previous result \cite{harissopulos2005} for most alpha energies (see Figure \ref{fig:xs-13C}). With the cross-section from Ref. \cite{harissopulos2005} the neutron yield from carbon and carbon containing materials is higher and agrees better with the neutron yield measured in Ref. \cite{gorshkov1964b} while being higher by about 20\% than that obtained in Ref. \cite{fernandes}.
The biggest difference between our calculations and data reported in Ref.~\cite{fernandes} is for $^{238}$U decay chain in iron but, surprisingly, the agreement for $^{232}$Th decay chain is pretty good, pointing probably to the additional uncertainty in the reported values in Ref.~\cite{fernandes} due to the yield extrapolation method. Experimental errors for Fe have not been quoted in that paper. 

Similar calculations of neutron yields and energy spectra have been performed using the NeuCBOT code which has now an option of using either JENDL-5 or TENDL libraries~\cite{Gromov:2023iuh}. Neutron yields from GEANT4 and associated SaG4n code using two libraries JENDL-5 \cite{jendl} and TENDL-2017 \cite{tendl2017} were reported in Ref.~\cite{mendoza2019}. A comparison between calculations with the optimised SOURCES4 and these codes for a few materials was included in Ref.~\cite{whitepaper}.

\begin{table}[!ht]
\centering
\caption{Neutron yield from ($\alpha, n$) reactions originated in $^{\mathrm{nat}}$U and $^{232}$Th decay chains in several materials as calculated by the SOURCES4 code. Neutron yield is given as the number of neutrons per gram of material per second per ppb of U or Th concentration. Spontaneous fission of $^{238}$U gives a neutron yield of $1.35 \times 10^{-11}$ n/g/s/ppb in all materials.}
\resizebox{\textwidth}{!}{\begin{tabular}{|c|c|c|c|c|c|}
\hline
\textbf{Element} & \textbf{$^{\mathrm{nat}}$U} & \textbf{$^{232}$Th} & \textbf{Compound} & \textbf{$^{\mathrm{nat}}$U} & \textbf{$^{232}$Th} \\ \hline
Li & \(7.12 \times 10^{-10}\) & \(2.95 \times 10^{-10}\) & Al\(_2\)O\(_3\) & \(8.53 \times 10^{-11}\) & \(4.17 \times 10^{-11}\) \\ \hline
Be & \(8.38 \times 10^{-9}\) & \(2.79 \times 10^{-9}\) & BeO & \(3.08 \times 10^{-9}\) & \(1.03 \times 10^{-9}\) \\ \hline
B & \(1.99 \times 10^{-9}\) & \(6.14 \times 10^{-10}\) & C\(_2\)F\(_4\) & \(9.76 \times 10^{-10}\) & \(3.90 \times 10^{-10}\) \\ \hline
C & \(1.41 \times 10^{-11}\) & \(5.48 \times 10^{-12}\) & CaCO\(_3\) & \(7.28 \times 10^{-12}\) & \(2.89 \times 10^{-12}\) \\ \hline
N & \(5.80 \times 10^{-11}\) & \(3.23 \times 10^{-11}\) & CaF\(_2\) & \(6.63 \times 10^{-10}\) & \(2.75 \times 10^{-10}\) \\ \hline
Na & \(4.13 \times 10^{-10}\) & \(1.93 \times 10^{-10}\) & CH\(_2\) & \(1.71 \times 10^{-11}\) & \(7.04 \times 10^{-12}\) \\ \hline
Mg & \(2.03 \times 10^{-10}\) & \(7.67 \times 10^{-11}\) & H\(_2\)O & \(3.98 \times 10^{-12}\) & \(1.39 \times 10^{-12}\) \\ \hline
Al & \(1.67 \times 10^{-10}\) & \(8.25 \times 10^{-11}\) & H\(_3\)BO\(_3\) & \(3.38 \times 10^{-10}\) & \(9.64 \times 10^{-11}\) \\ \hline
Si & \(2.18 \times 10^{-11}\) & \(1.01 \times 10^{-11}\) & MgO & \(1.20 \times 10^{-10}\) & \(4.56 \times 10^{-11}\) \\ \hline
P & \(2.85 \times 10^{-11}\) & \(1.94 \times 10^{-11}\) & Na\(_2\)CO\(_3\) & \(1.58 \times 10^{-10}\) & \(7.78 \times 10^{-11}\) \\ \hline
Cl & \(8.08 \times 10^{-11}\) & \(4.36 \times 10^{-11}\) & NaCl & \(2.20 \times 10^{-10}\) & \(1.06 \times 10^{-10}\) \\ \hline
Ar & \(1.52 \times 10^{-10}\) & \(9.00 \times 10^{-11}\) & NaF & \(8.16 \times 10^{-10}\) & \(3.50 \times 10^{-10}\) \\ \hline
Ca & \(1.80 \times 10^{-12}\) & \(1.22 \times 10^{-12}\) & PbF\(_2\) & \(4.39 \times 10^{-10}\) & \(1.74 \times 10^{-10}\) \\ \hline
Ti & \(3.63 \times 10^{-11}\) & \(3.18 \times 10^{-11}\) & SiO\(_2\) & \(1.41 \times 10^{-11}\) & \(5.98 \times 10^{-12}\) \\ \hline
Cr & \(1.40 \times 10^{-11}\) & \(1.45 \times 10^{-11}\) & \multirow{3}{*}{\shortstack{Stainless steel \\ Fe(66\%), Cr(17\%), Ni(12\%), \\ Mn(2\%), Mo(2\%), Si(1\%)}} & \multirow{3}{*}{\(7.38 \times 10^{-12}\)} & \multirow{3}{*}{\(8.91 \times 10^{-12}\)} \\ \cline{1-3}
Mn & \(9.29 \times 10^{-12}\) & \(1.04 \times 10^{-11}\) & & & \\ \cline{1-3}
Fe & \(4.74 \times 10^{-12}\) & \(6.68 \times 10^{-12}\) & & & \\ \hline
Ni & \(1.02 \times 10^{-13}\) & \(2.63 \times 10^{-13}\) & UC & \(2.50 \times 10^{-12}\) & \(1.02 \times 10^{-12}\) \\ \hline
Cu & \(3.67 \times 10^{-13}\) & \(1.07 \times 10^{-12}\) & UO\(_2\) & \(2.41 \times 10^{-12}\) & \(8.37 \times 10^{-13}\) \\ \hline
\end{tabular}}
\label{table:Yn_data}
\end{table}

\section{Conclusions} \label{sec:conclusions}

The optimised approach for calculating neutron yields using SOURCES4 code shows a good agreement with experimental data, proving the importance of using reliable cross-section data where available. In cases where experimental data are not available for a specific isotope, nuclear reaction codes such as TALYS, EMPIRE, or databases like JENDL provide a good alternative. The choice of models for the cross-sections is confirmed by comparing calculated neutron yields with alpha beam data and measurements of neutron yields from radioactive decay chains to ensure accuracy and reliability.
This approach ensures that the most accurate and recent information is used for neutron production calculations, which is very important for applications in rare event physics experiments. By increasing the number of discrete nuclear levels considered in TALYS 1.96, the modified code now accounts for all possible excited states for the final state nuclei, as available in RIPL-3.

The modified libraries of cross-sections and branching ratios used in this work are available at \cite{sources4tapes}. The full SOURCES4 source code cannot be shared due to licensing restrictions but the users who obtained the code from the Nuclear Energy Agency \cite{sources4c} are welcome to request the updated version from mihaela.parvu@unibuc.ro or v.kudryavtsev@sheffield.ac.uk.

\section{Acknowledgements}

V.K. and P.K. would like to thank the Science and Technology Facilities Council (STFC, UK, ST/W000547/1, ST/S003398/1, ST/Y004841/1) and the University of Sheffield for financial support. For M.P. this work was performed with the financial support of the Romanian Program PNCDI III, Programme 5, Module 5.2 CERN-RO 04/2022 and CERN-RO/CDI/2024\_001.

\bibliographystyle{elsarticle-num}
\bibliography{References}

\end{document}